\documentstyle[11pt,amssymb,amsfonts,amsthm,amssymb,amscd,amstext,epsfig]{article}

\def\ben{\begin{equation}}
\def\een{\end{equation}}

\let\a=\alpha \let\b=\beta  \let\d=\delta 
  \let\q=\theta 
\let\l=\lambda \let\m=\mu \let\n=\nu   
\let\s=\sigma   

  \let\D=\Delta

\let\na=\nabla
\let\pa=\partial
\def\be{\begin{equation}}
\def\ee{\end{equation}}
\def\ba{\begin{array}}
\def\ea{\end{array}}

\def\dalemb#1#2{{\vbox{\hrule height .#2pt
        \hbox{\vrule width.#2pt height#1pt \kern#1pt
                \vrule width.#2pt}
        \hrule height.#2pt}}}

\newcommand{\bea}{\begin{eqnarray}}
\newcommand{\eea}{\end{eqnarray}}

\begin{document}
\begin{flushright}
\hfill{DAMTP-2004-34} \\
{hep-th/0403281}
\end{flushright}

\begin{center}
\vspace{1cm} { \Large {\bf Casimir effect and thermodynamics of horizon instabilities}}

\vspace{1.5cm}

Sean A. Hartnoll

\vspace{0.3cm}

s.a.hartnoll@damtp.cam.ac.uk

\vspace{0.8cm}

{\it DAMTP, Centre for Mathematical Sciences,
 Cambridge University\\ Wilberforce Road, Cambridge CB3 OWA, UK}

\vspace{2cm}

\end{center}

\begin{abstract}

We propose a dual thermodynamic description of a classical
instability of generalised black hole spacetimes. From a
thermodynamic perspective, the instability is due to negative
compressibility in regions where the Casimir pressure is large.
The argument indicates how the correspondence between
thermodynamic and classical instability for horizons may be
extended to cases without translational invariance.

\end{abstract}

\pagebreak
\setcounter{page}{1}

\section{Introduction}

The intriguing connection between classical and local
thermodynamic instability
\cite{Gubser:2000ec,Gubser:2000mm,Reall:2001ag,Gregory:2001bd}
has partially been
behind a renewed interest in the Gregory-Laflamme instability of
black branes
\cite{Gregory:1994bj,Gregory:vy,Horowitz:2001cz,Gubser:2001ac,
Kol:2002xz,Wiseman:2002ti,Wiseman:2002zc,Kol:2003ja,Gubser:2002yi,
Hirayama:2001bi,Kang:2004hm}.
The connection states that translationally invariant horizons are
classically stable if and only if they are locally
thermodynamically stable
\cite{Gubser:2000ec,Gubser:2000mm,Reall:2001ag}. The objective of
this work is to suggest a dual thermodynamic description of a
recently discovered black hole instability. The novel aspects of
the connection presented here are that the horizon is
inhomogeneous and that the thermodynamic instability is due to a
Casimir effect.

Local thermodynamic stability requires the Helmholtz free energy
of the system to be a concave function of temperature and a convex
function of volume. The known instabilities of black branes are
due to a negative heat capacity, $C_V = - T
\pa^2 F / \pa T^2 |_V < 0$. In contrast, the instability in this
paper will be due to a negative isothermal compressibility
$K_T^{-1} = V \pa^2 F / \pa V^2 |_T < 0$.

We work within the context of the AdS/CFT correspondence
\cite{Maldacena:1997re,Gubser:1998bc,Witten:1998qj}. Amongst other
things, the AdS/CFT correspondence relates the physics of black
holes in AdS$_{d+2}$ to a thermal field theory living on the
`boundary' $S^1 \times S^d$ \cite{Witten:1998qj,Witten:1998zw}. In
the conclusion we note that it should be possible to extend the
ideas presented here to general inhomogeneous horizons, not
necessarily embedded in an AdS-like geometry.

Section 2 reviews an instability of generalised black holes
\cite{Gibbons:2002pq,Gibbons:2002th} with a negative cosmological
constant \cite{Hartnoll:2003as}. We further recall how the
criterion for classical instability of the black hole translates
into dual field theory language.

Section 3 calculates the energy momentum tensor of the thermal
field theory dual to the black hole spacetime using holographic
renormalisation
\cite{deHaro:2000xn,Balasubramanian:1999re,Kraus:1999di,Skenderis:2000in,Myers:1999ps}.
The thermal field theory has an ultrastatic background
$S^1\times{\mathcal{M}}$, with ${\mathcal{M}}$ an arbitrary
dimension $d$ Einstein manifold. There exists an ambiguity in the
energy momentum tensor due to local curvature polynomials of mass
dimension $(d+1)/2$. Generically, these produce an inhomogeneous
Casimir energy and pressure. We show how these terms may cause a
thermodynamic instability. This thermodynamic instability appears
to be a suitable dual description to the aforementioned classical
instability of the black hole bulk spacetime.

Section 4 illustrates the previous considerations in the case when
the horizon of the black hole is given by a B\"ohm metric on $S^5$
\cite{bohm,Gibbons:2002th}. The thermal field theory in this case
has background $S^1 \times S^5_{\text{B\"ohm}}$.

The last section is the discussion. The argument we have presented
is not rigorous and we clarify what would be needed for a
watertight discussion. Further applications of the Casimir effect
to horizon instabilities are suggested.

\section{Instability of generalised AdS black holes}

We start by reviewing the stability of generalised black holes
\cite{Gibbons:2002pq,Hartnoll:2003as} with a negative cosmological
constant. Generalised black holes take the form
\be\label{eq:metric}
ds^2 = -f(r) dt^2 + \frac{dr^2}{f(r)} + r^2 ds^2_d \,,
\ee
where $f(r) = 1 - (1+r_+^2/L^2)(r_+/r)^{d-1} + r^2/L^2$. The event horizon is at
$r=r_+$. If $ds^2_d$ were the round metric on $S^d$, then the spacetime would
be the usual Schwarzschild black hole in AdS space. However, the $d+2$ dimensional
vacuum Einstein equations allow the horizon metric $ds^2_d$ to be any $d$ dimensional Einstein metric
on a Riemannian manifold ${\mathcal{M}}$
\be\label{eq:normalise}
R_{a b} = (d-1) g_{a b} \,.
\ee
The resulting spacetime is called a generalised black hole. The asymptotic
geometry of the background has now changed and the dual thermal field theory
lives on $S^1\times{\mathcal{M}}$.

The stability of generalised black holes has recently been
investigated \cite{Hartnoll:2003as,Gibbons:2002pq,Gibbons:2002th}.
For large, $r_+/L \gg 1$, generalised AdS black holes one finds
\be\label{eq:stability}
\l_L < - \frac{r_+^2}{L^2} \times {\mathcal{O}}(1) \quad \Leftrightarrow
\quad \text{instability} \,,
\ee
where $\l_L$ is the minimum eigenvalue of the Lichnerowicz
operator on ${\mathcal{M}}$. Recall that the Lichnerowicz operator
acts on symmetric rank two tensors as
\be
(\D_L h)_{ab} = 2 R^c{}_{abd} h^d{}_c + R_{ca} h^c{}_b + R_{cb}
h^c{}_a - \nabla^c \nabla_c h_{ab} \, .
\ee
Note from (\ref{eq:stability}) that the minimum Lichnerowicz
eigenvalue needs to be very negative in order for instability to
occur. Examples of Einstein manifolds with large negative
Lichnerowicz eigenvalues are the B\"ohm metrics on $S^5 \ldots
S^9$ and on products of spheres \cite{bohm,Gibbons:2002th}. We
will use these metrics as examples below.

The criterion may be translated into field theory language using
the standard AdS/CFT dictionary to give \cite{Hartnoll:2003as}
\be
T^2 < -\l_L \times {\mathcal{O}}(1) \quad \Leftrightarrow
\quad \text{instability} \,,
\ee
where $T$ is the temperature of the field theory. Thus, the
AdS/CFT correspondence predicts a critical temperature in the dual
field theory, $T_C^2 \sim \mid \lambda_L \mid$.

The Lichnerowicz operator is not a particularly natural object to
consider in field theories where gravity is nondynamical.
On Einstein manifolds with positive curvature the minimum Lichnerowicz
eigenvalue is related to the Weyl curvature. Roughly, one expects \cite{Gibbons:2002th}
\be
\l_L^2 \sim C_{a b c d} C^{a b c d} |_{\text{max.}} \sim {\frak{R}}^2 \,,
\ee
where we use ${\frak{R}}$ to denote a typical magnitude of
curvature of the geometry ${\mathcal{M}}$, and $C_{a b c d} C^{a b
c d} |_{\text{max.}}$ is the maximum value taken by $C_{a b c d}
C^{a b c d}$ on the manifold. Putting these statements together,
the critical temperature becomes
\be\label{eq:critical}
T^2_C \sim {\frak{R}} \, .
\ee
Note that the Ricci scalar, $R$, is fixed, so in the limit of
large negative $\l_L$ we have $R \ll {\frak{R}}$. This hierarchy
is important for consistency because the large black hole limit
$r_+/L \gg 1$ translates into the field theory condition $T \gg
R$.

In the next section we present a potential field theory
instability occurring at this critical temperature.

\section{Casimir instability of field theory thermodynamics}

\subsection{Holographic energy momentum tensor}

The thermodynamics of the dual field theory in the strong coupling
regime may be studied using holographic renormalisation
\cite{deHaro:2000xn,Balasubramanian:1999re,Kraus:1999di,Skenderis:2000in,Myers:1999ps}.
This is the regime dual to the weakly curved gravitational description
in the bulk and is therefore the regime in which we expect to find a
field theory instability corresponding to the black hole instability.

For concreteness we will work now with seven dimensional black
holes and a six dimensional boundary. This allows the problem to
be embedded into the $AdS_7\times S^4$ version of the AdS/CFT
correspondence and will allow us in the next section to use five
dimensional B\"ohm metrics as concrete examples. We will work from
now on with a Euclidean signature, to describe finite temperature
physics.

The expectation value of the  energy momentum tensor of the dual
theory is given by
\be
< T^{\mu \nu} > = \left. \frac{2}{\sqrt{\det g}} \frac{\pa S}{\pa g_{\mu \nu}} \right|_{\text{renormalised}} \, ,
\ee
where $g_{\mu \nu}$ is the metric on the boundary of spacetime.
The gravitational action, $S$, has three parts: the
Einstein-Hilbert action, the Gibbons-Hawking boundary term and the
counterterm action. The counterterm action is defined on the
boundary of spacetime but differs crucially from the
Gibbons-Hawking term in that it is a local functional of intrinsic
boundary curvatures only. This implies that it does not affect the
bulk equations of motion. The counterterms are added to cancel
divergences in the action due to the diverging volume of the
spacetime at infinity.

The calculation of the dual energy momentum tensor for a general
spacetime with negative cosmological constant and a six
dimensional boundary has been performed, for example, in
\cite{deHaro:2000xn}. It is straightforward to apply their
formulae to the generalised black hole spacetime
(\ref{eq:metric}). Write the boundary metric as
\be\label{eq:themetric}
g_{\mu \nu}dx^{\mu}dx^{\nu} = d\tau^2 + g_{a b} \, dx^a dx^b
\ee
We have $\mu, \nu$ running over all the boundary coordinates and
$a,b\ldots$ running over the spatial directions of the boundary.
We find
\bea\label{eq:one}
<T_{0 0}> & = & \frac{3 L^5}{8\pi G_7 r_0^6}\left[-\frac{5}{6}
  \frac{r_+^4}{L^4}\left( 1 + \frac{r_+^2}{L^2} \right) + \frac{5}{48}
  \right] g_{0 0} + \left. T_{0 0} \right|_\text{f.c.} \nonumber \,, \\
<T_{a b}> & = & \frac{3 L^5}{8\pi G_7 r_0^6}\left[\frac{1}{6}
  \frac{r_+^4}{L^4}\left( 1 + \frac{r_+^2}{L^2} \right) - \frac{1}{48}
  \right] g_{a b} + \left. T_{a b} \right|_\text{f.c.} \,,
\eea
where f.c. stands for `finite counterterms', described below, and
$r_0$ is an arbitrary scale introduced during the renormalisation.
Dependence on $r_0$ will disappear shortly. The result
(\ref{eq:one}) may be reexpressed in terms of field theory
quantities by relating the radius of a large black hole, $r_+/L
\gg 1$ , to the field theory temperature, using the $AdS_7
\times S^4$ relation between the AdS length and Newton's constant
and introducing the Ricci scalar of $g_{ab}$:
\be
T \approx \frac{3}{2\pi} \frac{r_+}{r_0 L} \,, \quad \quad
\frac{1}{G_7} = \frac{2^4}{3\pi^2} \frac{N^3}{L^5} \,, \quad \quad
R = \frac{20}{r_0^2} \,.
\ee
This gives
\bea\label{eq:two}
<T_{0 0}> & = & - \frac{5 N^3}{3 \pi^3}
\left[\left(\frac{2\pi}{3}\right)^6 T^6 -
  \left(\frac{1}{5\cdot 2^3}\right)^3 R^3  \right] g_{0 0}  +
\left. T_{0 0} \right|_\text{f.c.} \nonumber \,, \\
<T_{a b}> & = &  \frac{N^3}{3 \pi^3}
\left[\left(\frac{2\pi}{3}\right)^6 T^6 -
  \left(\frac{1}{5\cdot 2^3}\right)^3 R^3  \right] g_{a b} +
\left. T_{a b} \right|_\text{f.c.} \,.
\eea
The Ricci scalar term is in fact negligible because the large
black hole limit implies $T^2 \gg R$.

The origin of the finite counterterm contribution is as follows.
The counterterm boundary action consists of curvature scalars. The
terms in the action that are scalars of mass dimension less than
six diverge as the boundary is taken to infinity and are used to
cancel the divergences of the bulk gravitational action. Scalars
of mass dimension greater than six are negligible as the boundary
goes to infinity. Scalars of mass dimension precisely six remain
finite. Thus, one is free to add these scalars to the countertem
action.

The same ambiguity also exists in the dual field theory. The
coefficient of a mass dimension six curvature scalar in a six
dimensional action is dimensionless and should therefore be viewed
as a renormalisable coupling. The value of the coupling is thus
generically not zero. Its value is not determined by the theory
but needs to be observed `experimentally'. The contribution of
these terms to the action might be thought of as a position
dependent cosmological constant. The variation of these terms with
respect to the boundary metric gives the contribution to the
energy momentum tensor that we have included above in
(\ref{eq:one}) and (\ref{eq:two}). Due to the fact that there this
constitutes a temperature independent contribution to the energy
and pressure, we call the presence of these terms a Casimir
effect.

\subsection{Thermodynamic instabilities - general argument}

Before considering explicitly the possible form of the finite
counterterm contribution to the energy momentum tensor, we will see how such
terms can produce thermodynamic instabilities. The following section
offers a concrete example. Recall that we are interested in manifolds
with a curvature scale much larger than the constant Ricci scalar ${\frak{R}}
\gg R$. The picture one has in mind is the following.
\begin{figure}[h]
\begin{center}
\vspace{4mm}
\epsfig{file=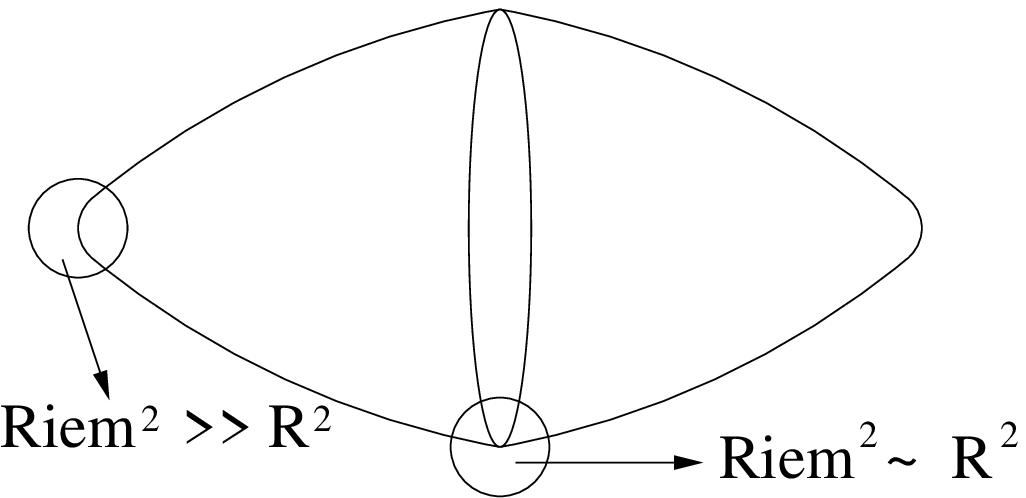,width=5cm}
\end{center}
\noindent {\bf Figure 1:} Manifold with two curvature scales,
$\text{Riem}^2 |_{\text{max}} \sim {\frak{R}}^2
\gg R^2$.
\end{figure}

In the region where the curvature scale ${\frak{R}}$ is large,
${\frak{R}} \gg R$, the pressure in (\ref{eq:two}) will be
\be\label{eq:pressure}
P \sim N^3 \left[ T^6 \pm {\frak{R}}^3 \right] \,.
\ee
We are neglecting positive order one coefficients. We have taken the
$N^3$ term as an overall factor because one would expect the Casimir
contribution to be proportional to the number of degrees of freedom.

If the temperature is sufficiently high, $T^2 \gg {\frak{R}}$,
then the field theory thermodynamics is just that of black body
radiation of ${\mathcal{O}}(N^3)$ species. However, if $T^2 <
{\frak{R}}$, then the Casimir pressure is dominant in the regions
of large curvature. This is consistent with the large temperature
approximation we are making, $T^2 \gg R$, because of the hierarchy
of curvatures ${\frak{R}} \gg R$. Thus, the qualitative
thermodynamic behaviour in the strong curvature regions appears to
change at around $T_C^2 \sim {\frak{R}}$.

The curvature ${\frak{R}}$ has an associated lengthscale
$L_{\frak{R}} = {\frak{R}}^{-1/2}$. This lengthscale gives the
size of the strong curvature region. The lengthscale associated
with the wavelength of thermal excitations is $L_T = T^{-1}$. In
the lower temperature regime $T^2 < T_C^2 \sim {\frak{R}}$, the
thermal wavelength is larger than the size of the strong curvature
region: $L_T > L_{\frak{R}}$. Therefore the bulk of the thermal
spectrum is not excited in this region. Effectively, the region is
not in thermodynamic equilibrium with the rest of the radiation.
To first approximation, the strongly curved region is at zero
temperature. In this case, let us consider the behaviour of the
strong curvature region as an independent system. The volume of
this region is
\be
V \sim L_{\frak{R}}^5 = \left(\frac{1}{{\frak{R}}}\right)^{5/2} \,.
\ee
Writing the pressure (\ref{eq:pressure}) in terms of the volume of
this region
\be
P \sim \pm \left(\frac{1}{V}\right)^{6/5} \,.
\ee
Consider the isothermal compressibility
\be
K_T^{-1} = - V \left(\frac{\pa P}{\pa V}\right)_T = \pm
\frac{6}{5}
\left(\frac{1}{V}\right)^{6/5}
\ee
The compressibility may also be expressed in terms of the Helmholtz
free energy by using the Maxwell relation $P = - \pa F / \pa V |_T$
\be
K_T^{-1} = V \left(\frac{\pa^2 F}{\pa V^2}\right)_T \,.
\ee
Thus where the compressibility is negative, the free energy is not
a convex function of the volume and the system is
thermodynamically unstable.

We see that there will be a thermodynamic instability at $T^2 =
T_C^2
\sim {\frak{R}}$ if the sign of the Casimir pressure in
(\ref{eq:pressure}) is negative. We will see below that in
practice the contributions to the pressure come with both positive
and negative signs. The present argument indicates the generic
possibility of instability, rather than a necessary instability.
The localisation of the instability to the strong curvature
regions is mirrored by the classical black hole instability, where
the unstable mode is concentrated in these regions
\cite{Gibbons:2002th}.

The dual operator to the unstable mode \cite{Hartnoll:2003as}
is a linear combination of
spatial components of the energy momentum tensor. Below the critical
temperature one expects the vacuum expectation value of this
operator, which are components of the pressure, to shift. This
suggests that there is a new stable phase where the curvature does not
give the dominant contribution to the pressure.

This instability should be compared with previous matchings
between dynamical and thermodynamic instability
\cite{Gubser:2000ec,Gubser:2000mm,Reall:2001ag,Gregory:2001bd}. The instability
in these works is due to a negative heat capacity
\be
C_V = T \left(\frac{\pa S}{\pa T}\right)_V = - T \left(
\frac{\pa^2
  F}{\pa T^2} \right)_V \,.
\ee

\subsection{Contributions to the Casimir pressure}

We can be more precise about the curvature terms contributing to
the energy-momentum tensor in the strongly curved regions. In six
spacetime dimensions, there are 17 curvature scalars that have
mass dimension six \cite{Fulling:vm}. These will generically be
present in the action and their variation will contribute to the
energy momentum tensor. However, not all of the contributions will be
large. The spatial background is Einstein, and consequently has constant
Ricci scalar, $R$, and Ricci tensor. Some of the tensors will be
${\mathcal{O}}(R^3)$ or ${\mathcal{O}}(R^2 {\frak{R}})$ or
${\mathcal{O}}(R{\frak{R}}^2)$ instead of
${\mathcal{O}}({\frak{R}}^3)$. These will be negligible in the
region where ${\frak{R}} \gg R$.

Given that the field theory does not specify the coefficients of
the curvature terms in the action, there is not much to be gained
at this stage from starting with the action. Instead of
calculating the variation of the 17 curvature scalars, we will
start by directly considering all possible rank two tensors that
have mass dimension six. Although there are many such tensors
\cite{Fulling:vm}, if we consider a background of the form
$S^1\times{\mathcal{M}}$, with ${\mathcal{M}}$ Einstein, then
there are only eight independent rank two tensors with mass
dimension six that are ${\mathcal{O}}({\frak{R}}^3)$. Explicitly,
\be
\left. T_{\m \n} \right|_\text{f.c.} = \sum_{i=1}^8 k_i K_{\m
 \n}^{(i)} + {\mathcal{O}}(R{\frak{R}}^2)\,,
\ee
for some coefficients $k_i$ and where the tensors $K_{\m
\n}^{(i)}$ are\footnote{For a general six dimensional metric there
is a ninth independent tensor $R^{pqrs} R_{pqtu}
R_{rs}{}^{tu}g_{\m \n}$, but this is not independent in five
dimensions, or in the case $S^1\times{\mathcal{M}}$. Thus in our
backgrounds it is not ${\mathcal{O}}({\frak{R}}^3)$.}
\bea\label{eq:tensors}
K_{\m\n}^{(1) \cdots (8)} & = & \left\{ R^{pqrs} \na_{(\m}
\na_{\n)} R_{pqrs}
\, , \; \na_{\m} R^{pqrs}
\na_\n R_{pqrs} \, , \; \na_s R^{pqr}{}_\m \na^s R_{pqr\n} , \;
R^{pqrs} R_{pq}{}^t{}_\m R_{rst\n} \,, \right. \nonumber \\
 & & \left. R^{pqrs} R_p{}^t{}_{r\m} R_{qts\n} \, , \; R^{pqrs}
R_{pqr}{}^t R_{s\m t\n} \, , \; \na_t R^{pqrs} \na^t R_{pqrs}
g_{\m\n}\, ,\; R^{pqrs}R_p{}^t{}_r{}^u R_{qtsu} g_{\m\n}
\right\} \, .\nonumber \\
\eea
All other linearly independent rank two tensors with mass
dimension six either vanish or contain factors of the Ricci scalar
when considered in an Einstein background. This may be worked out
from the basis of tensors given in \cite{Fulling:vm}. The $S^1$
factor in the metric does not contribute to any of the curvature
tensors and so all the summations involving curvature tensors are
over the five spatial directions - compare with the notation of
equation (\ref{eq:themetric}).

Not all of these terms are obtainable from varying an action. We
must impose two constraints. Firstly, diffeomorphism invariance of
the action implies that the energy momentum tensor is transverse
(the Noether constraint) because under the diffeomorphism $\d
g_{\m\n} = \na_{(\m}\xi_{\n)}$:
\be
0 = \d_{\xi}S = \int \frac{\d S}{\d g_{\m \n}} \na_{(\m}\xi_{\n)}
d\Omega = - \int \left(\na_\m \frac{\d S}{\d g_{\m
\n}}\right)\xi_{\n} d\Omega \quad \Rightarrow \quad \na_\m
\frac{\d S}{\d g_{\m\n}} = \na_\m T^{\m \n} = 0 \,.
\ee
Secondly, in six dimensions an action given by scalars with mass
dimension six is invariant under constant Weyl rescalings. This
implies that the trace of the corresponding energy momentum tensor
must be a total derivative. Under $\d g_{\m \n} = \s g_{\m \n}$:
\be
\int \frac{\d S}{\d g_{\m \n}} \s g_{\m\n} d\Omega =
\d_{\s}S = \int \na_\m \s J^\m d\Omega = - \int \s \na_\m J^\m d\Omega
\quad \Rightarrow \quad g_{\m \n} \frac{\d S}{\d g_{\m \n}} = T = - \na_\m J^\m \,,
\ee
for some vector $J^\m$. This total derivative contributes to the
non-universal part of the Weyl anomaly \cite{Bastianelli:2000rs,Deser:yx}.

The trace constraint is the easier to apply. The possible
candidates for $J^\m$ are rank one tensors with mass dimension
five or indeed zero. The only relevant nonzero possibility is
$J_\m = R^{pqrs} \na_\m R_{pqrs}$ \cite{Fulling:vm}. Considering
the eight tensors in (\ref{eq:tensors}) it follows that there are
seven admissible combinations
\be\label{eq:tracefree}
\left\{ K^{(1)}_{\m \n} + K^{(2)}_{\m \n}\,, \;  K^{(1)}_{\m \n} +
K^{(3)}_{\m \n}\,, \;  K^{(1)}_{\m \n} + \frac{1}{6} K^{(7)}_{\m
\n} \,, \; K^{(5)}_{\m \n} -
\frac{1}{6} K^{(8)}_{\m \n} \,, \; K^{(6)}_{\m \n} \,, \; K^{(4)}_{\m \n} \,, \;
K^{(1)}_{\m \n} + 4 K^{(5)}_{\m \n} \right\} \,,
\ee
where, in taking the trace, any terms arising that are not
${\mathcal{O}}({\frak{R}}^3)$ are ignored. It is assumed that any
contributions to the trace smaller than
${\mathcal{O}}({\frak{R}}^3)$ may be cancelled using other small
terms. The footnote of the previous page is important for this
calculation.

Now consider the divergences of these seven tensors. We need
combinations of the tensors in (\ref{eq:tracefree}) that have
vanishing divergence. The appendix indicates how one goes about
finding the combinations systematically. The result is four linearly
independent tensors
\bea\label{eq:divfree}
H^{(1)}_{\m\n} & = & K^{(4)}_{\m\n} -2 K^{(5)}_{\m\n} - K^{(6)}_{\m\n} + \frac{1}{3}K^{(8)}_{\m\n}  \,, \nonumber \\
H^{(2)}_{\m\n} & = &  - K^{(1)}_{\m\n} + K^{(2)}_{\m\n}
-16 K^{(5)}_{\m\n} + \frac{4}{3} K^{(8)}_{\m\n}  \,, \nonumber \\
H^{(3)}_{\m\n} & = & -\frac{1}{2} K^{(1)}_{\m\n} + K^{(3)}_{\m\n} -10 K^{(5)}_{\m\n}
+ \frac{2}{3} K^{(8)}_{\m\n} \,, \nonumber \\
H^{(4)}_{\m\n} & = & - K^{(1)}_{\m\n} -8 K^{(5)}_{\m\n} + \frac{1}{2} K^{(7)}_{\m\n}
- \frac{4}{3} K^{(8)}_{\m\n} \,.
\eea

Let us summarise what we have just achieved. We wanted to
calculate possible contributions to the energy momentum tensor
coming from a mass dimension six curvature scalar in the action.
Further, we were only interested in terms that are
${\mathcal{O}}({\frak{R}}^3)$ and hence give the dominant
contribution in the strongly curved regions. These contributions
had to be rank two tensors with mass dimension six. We have shown
that the only tensors satisfying all these conditions are the four
in equation (\ref{eq:divfree}). We may now calculate these terms
in a concrete background.

\section{Dimension six tensors of B\"ohm metrics on $S^5$}

The B\"ohm metrics \cite{bohm} serve as an illustration of the
ideas we have introduced. We will consider B\"ohm metrics on
$S^5$. The metrics have the form
\be
ds^2_5 = d\q^2 + a(\q)^2 d\Omega^2_2 + b(\q)^2
d\widetilde{\Omega}^2_2 \,,
\ee
where $d\Omega^2_2$ and $d\widetilde{\Omega}^2_2$ are round
metrics on spheres $S^2$ and $\widetilde{S}^2$, with polar coordinates
$(\a,\b)$ and $(\tilde{\a},\tilde{\b})$ respectively. The
coordinate $\q$ has range $0$ to $\q_{\text{f}}$. The Einstein
equations imply some nonlinear differential equations for $a(\q)$
and $b(\q)$. The boundary conditions for topology $S^5$ are that
$a(0)=0, b(0)=b_0$ and $a(\q_{\text{f}})=a_0, b(\q_{\text{f}})=0$.
In fact, the solutions to the equations on $S^5$ have $a(\q) =
b(\q_{\text{f}}-\q)$. There is a discrete infinity of solutions,
at specific values of $b_0$. Solutions exist for arbitrarily small
$b_0$. As $b_0 \to 0$ the metric develops conical singularities at
$\q=0,\q=\q_{\text{f}}$. Thus $b_0$ gives the resolution of the
singularity. We expect to find ${\frak{R}} \sim b_0^{-2}$.

We can evaluate the four contributions to the energy momentum
tensor of (\ref{eq:divfree}) at the points $\theta=0$ and
$\q=\q_{\text{f}}$ using Taylor series expansions for the
functions $a(\q)$ and $b(\q)$ \cite{Gibbons:2002th}. The results
for $b_0
\ll 1$ are contained in Table 1 below.
One may obtain the values at $\q=\q_{\text{f}}$ by letting
$H^{(i)}_\a{}^\a \leftrightarrow H^{(i)}_{\tilde{\a}}{}^{\tilde{\a}} $,
$H^{(i)}_\b{}^\b \leftrightarrow H^{(i)}_{\tilde{\b}}{}^{\tilde{\b}}$
and $H^{(i)}_\q{}^\q$ remains the same.

\begin{table}[h]
\begin{center}

  \begin{tabular}{|c||c|} \hline
Tensor components & At $\q = 0$  \\ \hline \hline
\rule[-5mm]{0mm}{13mm} $H^{(1)}_\q{}^\q = H^{(1)}_\a{}^\a =
H^{(1)}_\b{}^\b = H^{(1)}_{\tilde{\a}}{}^{\tilde{\a}} =
H^{(1)}_{\tilde{\b}}{}^{\tilde{\b}}$ & $\displaystyle \frac{32}{27 b_0^6}$  \\ \hline

\rule[-5mm]{0mm}{13mm} $\left\{ H^{(2)}_\q{}^\q ,
H^{(2)}_\a{}^\a = H^{(2)}_\b{}^\b ,
H^{(2)}_{\tilde{\a}}{}^{\tilde{\a}} = H^{(2)}_{\tilde{\b}}{}^{\tilde{\b}} \right\}$ &
$\displaystyle \left\{ \frac{64}{27 b_0^6}, -\frac{128}{27 b_0^6}, -\frac{256}{27 b_0^6} \right\}$  \\ \hline

\rule[-5mm]{0mm}{13mm} $\left\{ H^{(3)}_\q{}^\q ,
H^{(3)}_\a{}^\a = H^{(3)}_\b{}^\b ,
H^{(3)}_{\tilde{\a}}{}^{\tilde{\a}} = H^{(3)}_{\tilde{\b}}{}^{\tilde{\b}} \right\}$ &
$\displaystyle \left\{
{\mathcal{O}}(\frac{1}{b_0^4}), - \frac{32}{9 b_0^6} , -\frac{176}{27b_0^6} \right\}$   \\ \hline

\rule[-5mm]{0mm}{13mm} $\left\{ H^{(4)}_\q{}^\q ,
H^{(4)}_\a{}^\a = H^{(4)}_\b{}^\b ,
H^{(4)}_{\tilde{\a}}{}^{\tilde{\a}} = H^{(4)}_{\tilde{\b}}{}^{\tilde{\b}} \right\}$ &
$\displaystyle \left\{
- \frac{64}{27 b_0^6} , -\frac{256}{27 b_0^6} , -\frac{320}{27 b_0^6}\right\}$   \\ \hline
  \end{tabular}

\vspace{0.3cm}
{\bf Table 1:} Contributions to the pressure at $\q=0$ for B\"ohm
backgrounds. $P^{(i)}_x = k_i H^{(i)}_x{}^x$.
\end{center}
\end{table}

We are working in units where the Ricci scalar $R=20$, so the
condition $b_0 \ll 1$ is the introduction of a new scale: $b_0^{-2}
\sim {\frak{R}} \gg R$. The first observation we can make about Table
1 is that the contribution to the pressure is
${\mathcal{O}}({\frak{R}}^3)$ as expected. The pressure in each
direction is given by $P^{(i)}_x = k_i H^{(i)}_x{}^x$. The second
observation is that contributions to the pressure from
$H^{(2)}_{\m\n}$ have different signs for different directions.
Thus, whichever the sign of the coefficient, $k_2$, of
$H^{(2)}_{\m\n}$ in the energy momentum tensor, one of these
pressures will be negative, contributing towards a thermodynamic
instability as outlined above. Of course, all the pressures could
still be made positive by adding a larger positive contribution
from one of the other tensors.

\section{Summary, discussion and open questions}

Large generalised black holes with a negative cosmological
constant tend to have dynamical instabilities if the horizon has a
curvature scale which in some region is large compared to the
constant Ricci scalar of the horizon, ${\frak{R}} \gg R$. This
predicts a critical temperature in the dual field theory: $T_C^2
\sim {\frak{R}}$.

We specialised to the case of a six dimensional field theory, as
this corresponds to the lowest dimension in which unstable black
holes are known and further may be considered within the $AdS_7
\times S^4$ version of the AdS/CFT duality.

We have argued that field theories on curved backgrounds with a
curvature scale ${\frak{R}} \gg R$ can generically have a
thermodynamic instability at $T_C^2 \sim {\frak{R}}$ due to the
strong curvature region having negative compressibility. The
instability is caused by contributions to the pressure from mass
dimension six curvature tensors. These come from mass dimension
curvature six scalars that are renormalisable vacuum terms in the
field theory action. We calculated the explicit form of the
contribution to the energy momentum tensor, and evaluated these
tensors for the case of B\"ohm backgrounds.

There are various unsatisfactory aspects of our treatment. We do
not have a precise matching. Whilst a precise criterion for black
hole instability is known \cite{Hartnoll:2003as}, this is not the
case in field theory. This is because the coefficients of the
curvature terms in the energy momentum tensor are not predicted by
the field theory, but are rather empirically determined, just like
coupling constants. Yet, the instability itself depends on the
total contribution to the pressure having a specific sign.

This situation potentially raises a conceptual problem for the
AdS/CFT correspondence. The field theory has undetermined
coefficients that correspond to finite terms in the boundary
action on the gravitational side of the duality. However, these
terms do not influence the dynamics of the bulk, and the classical
stability of the bulk in particular. A possible resolution of the
problem is that the bulk spacetime determines a specific
renormalisation scheme for the field theory terms in question. In
order for the matching of instabilities to be made precise, this
renormalisation scheme would have to be known explicitly.

Nonetheless, the qualitative agreement is suggestive. It fits well
into the recent discovery of agreement between thermodynamic and
classical instabilities of horizons, and extends the connection to
cases without translational invariance. An exciting possibility is
that the ideas presented here could be adapted for use in other
contexts where the stability of inhomogeneous horizons is of
interest. In more general cases the Casimir energy would
presumably not be that of a dual field theory, but rather due to
gravitational degrees of freedom. A possible application of
current interest is the stability of nonuniform black branes
\cite{Gubser:2001ac,Kol:2002xz, Wiseman:2002ti,Wiseman:2002zc}.
Can the stability be studied from a thermodynamic perspective? Can
one associate a Casimir energy and pressure with general
inhomogeneous horizons?

\section*{Acknowledgments}

During this work I've had helpful comments from Ian Drummond, Gary
Gibbons, Hugh Osborne, Ron Horgan, Rob Myers and Kostas Skenderis. I'd also
like to thank David Berman, Neil Constable, Dumitru Ghilancea,
Robert Helling, Giusseppe Policastro and Christian Stahn for some
interesting conversations. This work is financially supported by
the Sims scholarship.

\appendix

\section{Divergences of mass dimension six two-tensors}

This appendix contains formulae that are necessary for finding the
independent linear combinations of the tensors in
(\ref{eq:tracefree}) that have vanishing divergence. This is done by
expressing the divergences of each rank two tensor in terms of
independent rank one tensors. The independent rank one tensors will be
\bea
A^{(1)}_\n & = & \na_a R_{pqrs} \na_\n \na^a  R^{pqrs} \,, \nonumber \\
A^{(2)}_\n & = & R_{pqrs} R_e{}^q{}_{a\n} \na^a R^{pers} \,, \nonumber \\
A^{(3)}_\n & = & R_{pqrs} R_e{}^q{}_a{}^r \na_\n R^{epas} \,, \nonumber \\
A^{(4)}_\n & = & R_{pqrs} R_{ea}{}^q{}_\n \na^a R^{pers} \,.
\eea
With a little work, one may then show that
\be\label{eq:decomp}
\left. \begin{array}{lll}
\na^\m K^{(1)}_{\m\n} & = & A^{(1)}_\n - 4 A^{(2)}_\n + 8
A^{(3)}_\n  \\
\na^\m K^{(2)}_{\m\n} & = & A^{(1)}_\n + 4 A^{(2)}_\n + 4 A^{(3)}_\n \\
\na^\m K^{(3)}_{\m\n} & = & \frac{1}{2} A^{(1)}_\n + 3 A^{(2)}_\n
+A^{(3)}_\n  \\
\na^\m K^{(4)}_{\m\n} & = & 2 A^{(4)}_\n \\
\na^\m K^{(5)}_{\m\n} & = & \frac{1}{2} A^{(2)}_\n - \frac{1}{2} A^{(3)}_\n  \\
\na^\m K^{(6)}_{\m\n} & = & - A^{(2)}_\n + 2 A^{(4)}_\n    \\
\na^\m K^{(7)}_{\m\n} & = & 2 A^{(1)}_\n  \\
\na^\m K^{(8)}_{\m\n} & = &  -3 A^{(3)}_\n  \\
\end{array}
\right\} + {\mathcal{O}}(R{\frak{R}}^{5/2}) \,,
\ee
where the ${\mathcal{O}}(R{\frak{R}}^{5/2})$ indicates that
throughout the calculation, all Ricci tensors have been dropped.
The results are exact if $R_{\m\n} = 0$. Recall that we do this
because we are interested in the dominant contributions in the
region where ${\frak{R}} \gg R$.

There is a fifth tensor that appears in these calculations which one
might think is linearly independent
\be
A^{(5)}_\n = R_{pqrs} R^r{}_e{}^s{}_a \na_\n R^{pqae} \,.
\ee
However, this can be seen to be a multiple of $\nabla_\nu \left(R^{pqrs} R_{pqtu}
R_{rs}{}^{tu}\right)$. As we noted above, for the backgrounds we are
considering, $R^{pqrs} R_{pqtu}R_{rs}{}^{tu}$ is not
${\mathcal{O}}({\frak{R}}^3)$. We are only keeping track of leading
order terms, so this term is not linearly independent.

Given these expressions for the divergences in terms of linearly
independent one-tensors (\ref{eq:decomp}), it is simple linear algebra
to find the two divergence free tensors of equation (\ref{eq:divfree}).

\end{document}